\documentclass[12pt]{article}
\usepackage[latin9]{inputenc}
\usepackage[a4paper]{geometry}
\usepackage{float}
\usepackage{amsmath}
\usepackage{amssymb}
\usepackage{graphicx}
\usepackage[authoryear]{natbib}

\makeatletter

\providecommand{\tabularnewline}{\\}

\date{3 August, 2022}

\makeatother

\begin{document}
\title{Weak Instruments, First-Stage Heteroskedasticity, the Robust F-Test
and a GMM Estimator with the Weight Matrix Based on First-Stage Residuals}
\author{Frank Windmeijer\thanks{Email: frank.windmeijer@stats.ox.ac.uk} \\
 Department of Statistics and Nuffield College\\
 University of Oxford, UK}
\maketitle
\begin{abstract}
\baselineskip=13pt

\noindent This paper is concerned with the findings related to the
robust first-stage F-statistic in the Monte Carlo analysis of \citet{IAndrewsREStat2018},
who found in a heteroskedastic grouped-data design that even for very
large values of the robust F-statistic, the standard 2SLS confidence
intervals had large coverage distortions. This finding appears to
discredit the robust F-statistic as a test for underidentification.
However, it is shown here that large values of the robust F-statistic
do imply that there is first-stage information, but this may not be
utilized well by the 2SLS estimator, or the standard GMM estimator.
An estimator that corrects for this is a robust GMM estimator, denoted
GMMf, with the robust weight matrix not based on the structural residuals,
but on the first-stage residuals. For the grouped-data setting of
Andrews (2018), this GMMf estimator gives the weights to the group
specific estimators according to the group specific concentration
parameters in the same way as 2SLS does under homoskedasticity, which
is formally shown using weak instrument asymptotics. The GMMf estimator
is much better behaved than the 2SLS estimator in the \citet{IAndrewsREStat2018}
design, behaving well in terms of relative bias and Wald-test size
distortion at more standard values of the robust F-statistic. We show
that the same patterns can occur in a dynamic panel data model when
the error variance is heteroskedastic over time. We further derive
the conditions under which the Stock and Yogo (2005) weak instruments
critical values apply to the robust F-statistic in relation to the
behaviour of the GMMf estimator.
\end{abstract}


\thispagestyle{empty}

\noindent \baselineskip=20pt

\pagebreak{}

\pagenumbering{arabic} \setcounter{page}{1}

\section{Introduction}

\label{Sec:Intro}It is commonplace to report the first-stage F-statistic
as a test for underidentification in linear single endogenous variable
models estimated by two-stage least squares (2SLS). This could either
be a non-robust or robust version of the test, with robustness to
for example heteroskedasticity, serial correlation and/or clustering.
Under maintained assumptions, these are valid tests for the null $H_{0}:\pi=0$
in the first-stage linear specification $x=Z\pi+v$, where $x$ is
the endogenous explanatory variable in the model of interest $y=x\beta+u$,
and $Z$ are the instruments. If the null is not rejected, then this
is an indication that the relevance condition of the instruments does
not hold and that the 2SLS estimator does not provide a meaningful
estimate of the parameter of interest $\beta$. A rejection of the
null does, however, not necessarily imply that the 2SLS estimator
is well behaved. This follows the work of \citet{StaigerStock1997}
and \citet{StockYogo2005}, with the latter providing critical values
for the first-stage non-robust F-statistic for null hypotheses of
weak instruments in terms of bias of the 2SLS estimator relative to
that of the OLS estimator and Wald test size distortion. These non-robust
weak instruments F-tests are valid only under conditional homoskedasticity,
no serial correlation and no clustering of both the first-stage errors
$v$ and the structural errors $u$, and do not apply to the robust
F-test in general designs, see \citet{BunHaan2010}, \citet{OleaPflueger2013}
and \citet{IAndrewsREStat2018}. For general designs \citet{OleaPflueger2013}
proposed the effective first-stage F-statistic and critical values
linked to the Nagar bias of the 2SLS estimator, whereas \citet{IAndrewsREStat2018}
obtained valid two-step identification robust confidence sets.

This paper is concerned with the findings related to the robust F-statistic
in the Monte Carlo analysis of \citet[Supplementary Appendix (SA)]{IAndrewsREStat2018}.
In a cross sectional heteroskedastic design he found that even for
very large values of the robust F-statistic, the standard 2SLS confidence
intervals had large coverage distortions. For example, for a high
endogeneity design, ``the 2SLS confidence set has a 15\% coverage
distortion even when the mean of the first-stage robust F-statistic
is 100,000'', \citet[SA, p 11]{IAndrewsREStat2018}. This is a striking
finding and appears to discredit the robust F-statistic as a test
for underidentification. However, it is shown here that large values
of the robust F-statistic do imply that there is first-stage information,
but this may not be utilized well by the 2SLS estimator, or GMM estimators
that incorporate heteroskedasticity in the structural error $u$ only.

The \citet{IAndrewsREStat2018} design is the same as a grouped data
one, see \citet{AngristJoE1991} and the discussion in \citet{AngristPischke2009},
where the instruments are mutually exclusive group membership indicators.
Denoting the groups by $s=1,...,S$, the group specific concentration
parameter values are determined by the ratios $\pi_{s}^{2}/\sigma_{v,s}^{2}$,
where $\sigma_{v,s}^{2}$ is the group specific variance of the first-stage
error $v$. The 2SLS estimator is a weighted average of the group
specific estimators of $\beta$, giving more weight to large concentration
parameter groups if $v$ is homoskedastic. However, as shown in Section
\ref{Sec:Model}, this may not happen under heteroskedasticity, where
2SLS gives more weight to high variance $\sigma_{v,s}^{2}$ groups,
everything else constant. In the design of \citet{IAndrewsREStat2018}
we consider here, there is one informative group, leading to the large
value of the robust F-statistic, but this group has a small variance
$\sigma_{v,s}^{2}$, and therefore gets only a relatively small weight
in the 2SLS estimator.

An estimator that correctly gives larger weights to more informative
groups is a robust GMM estimator, not using the structural residuals
$\widehat{u}$, but the first-stage residuals $\widehat{v}$ in the
robust weight matrix. This estimator, called GMMf, is introduced in
Section \ref{Sec:AltGMM} and gives the weights to the group-specific
estimators according to the group-specific concentration parameters
in the same way as 2SLS does under homoskedasticity. This is further
formally shown using weak instrument asymptotics in Section \ref{Sec:WIA}.
Section \ref{Sec:varu} discusses the potential problems of the standard
GMM estimator that uses a robust weight matrix based on the conditional
variances of the structural errors $u$. Monte Carlo results in Section
\ref{Sec:MC} show that the GMMf estimator exploits the available
information well, with much better relative bias and Wald test size
properties than the 2SLS estimator for values of the robust F-statistic
in line with those of the non-robust F-statistic and behaviour of
the 2SLS estimator in the homoskedastic case.

Section \ref{Sec:Panel} shows that similar patterns can occur when
considering a simple AR(1) dynamic panel data model with heteroskedasticity
of the idiosyncratic shocks in the time dimension, with estimation
based on the forward orthogonal deviations transformation.

For a general setting, we report in Section \ref{Sec:wit} the conditions
under which the Stock and Yogo (2005) critical values can be applied
to the robust F-statistic in relation to the behaviour of the GMMf
estimator. These conditions are derived in Appendix \ref{Sec:AppWIA}.
Whilst these have limited applicability, the fully homoskedastic design
is a special case.

\section{Grouped-Data IV Model, First-Stage F-Statistic and 2SLS Weights }

\label{Sec:Model}

We consider the model as in \citet[SA C.3]{IAndrewsREStat2018}, which
is the same as a grouped-data IV setup,
\begin{eqnarray*}
y_{i} & = & x_{i}\beta+u_{i}\\
x_{i} & = & z_{i}^{\prime}\pi+v_{i},
\end{eqnarray*}
for $i=1,...,n$, where the $S$-vector $z_{i}\in\left\{ e_{1},...,e_{S}\right\} $,
with $e_{s}$ an $S$-vector with $s$th entry equal to $1$ and zeros
everywhere else, for $s=1,\ldots,S$. Assumptions for standard asymptotic
normality results for the IV estimator hold and the variance of the
limiting distribution of the parameters can be estimated consistently.

The variance-covariance structure for the errors is modeled fully
flexibly by group, and specified as
\[
\left(\left(\begin{array}{c}
u_{i}\\
v_{i}
\end{array}\right)|z_{i}=e_{s}\right)\sim\left(0,\Sigma_{s}\right).
\]
\begin{equation}
\Sigma_{s}=\left[\begin{array}{cc}
\sigma_{u,s}^{2} & \sigma_{uv,s}\\
\sigma_{uv,s} & \sigma_{v,s}^{2}
\end{array}\right].\label{eq:Sigmas}
\end{equation}
At the group level, we therefore have for group member $j$ in group
$s$
\begin{eqnarray}
y_{js} & = & x_{js}\beta+u_{js}\label{modyjs}\\
x_{js} & = & \pi_{s}+v_{js}\label{modxjs}
\end{eqnarray}
\[
\left(\begin{array}{c}
u_{js}\\
v_{js}
\end{array}\right)\sim\left(0,\Sigma_{s}\right)
\]
for $j=1,...,n_{s}$ and $s=1,...,S$, with $n_{s}$ the number of
observations in group $s$, $\sum_{s=1}^{S}n_{s}=n$, see also \citet{BekkervdPloeg2005}.
We assume that $\lim_{n\rightarrow\infty}\frac{n_{s}}{n}=f_{s}$,
with $0<f_{s}<1$.

The OLS estimator of $\pi_{s}$ is given by $\widehat{\pi}_{s}=\overline{x}_{s}=\frac{1}{n_{s}}\sum_{j=1}^{n_{s}}x_{js}$
and $Var\left(\widehat{\pi}_{s}\right)=\sigma_{v,s}^{2}/n_{s}$. The
OLS residual is $\widehat{v}_{js}=x_{js}-\overline{x}_{s}$ and the
estimator for the variance is given by $V\widehat{a}r\left(\widehat{\pi}_{s}\right)=\widehat{\sigma}_{v,s}^{2}/n_{s}$,
where $\widehat{\sigma}_{v,s}^{2}=\frac{1}{n_{s}}\sum_{j=1}^{n_{s}}\widehat{v}_{js}^{2}$.
Let $Z$ be the $n\times S$ matrix of instruments. For the vector
$\pi$ the OLS estimator is given by
\[
\widehat{\pi}=\left(Z^{\prime}Z\right)^{-1}Z^{\prime}x=\left(\overline{x}_{1},\overline{x}_{2},...,\overline{x}_{S}\right)^{\prime}.
\]
Let
\begin{eqnarray}
\widehat{\Omega}_{v} & = & \sum_{i=1}^{n}\widehat{v}_{i}^{2}z_{i}z_{i}^{\prime}\label{omev}\\
 & = & \text{diag}\left(n_{s}\widehat{\sigma}_{v,s}^{2}\right),\nonumber
\end{eqnarray}
where $\text{diag}\left(q_{s}\right)$ is a diagonal matrix with $s$th
diagonal element $q_{s}$. Then the robust estimator of $Var\left(\widehat{\pi}\right)$
is given by
\begin{eqnarray*}
V\widehat{a}r_{r}\left(\widehat{\pi}\right) & = & \left(Z^{\prime}Z\right)^{-1}\widehat{\Omega}_{v}\left(Z^{\prime}Z\right)^{-1}\\
 & = & \text{diag}\left(\widehat{\sigma}_{v,s}^{2}/n_{s}\right).
\end{eqnarray*}
The non-robust variance estimator is
\begin{eqnarray*}
V\widehat{a}r\left(\widehat{\pi}\right) & = & \left(\frac{1}{n}\sum_{i=1}^{n}\widehat{v}_{i}^{2}\right)\left(Z^{\prime}Z\right)^{-1}\\
 & = & \left(\sum_{s=1}^{S}\frac{n_{s}}{n}\widehat{\sigma}_{v,s}^{2}\right)\text{diag}\left(\frac{1}{n_{s}}\right).
\end{eqnarray*}

The group (or instrument) specific IV estimators for $\beta$ are
given by
\begin{equation}
\widehat{\beta}_{s}=\frac{z_{s}^{\prime}y}{z_{s}^{\prime}x}=\frac{\overline{y}_{s}}{\overline{x}_{s}},\label{betasind}
\end{equation}
with $\overline{y}_{s}=\frac{1}{n_{s}}\sum_{j=1}^{n_{s}}y_{js}$,
and the 2SLS estimator for $\beta$ is, with $P_{Z}=Z\left(Z^{\prime}Z\right)^{-1}Z^{\prime}$,
\begin{eqnarray*}
\widehat{\beta}_{2sls} & = & \left(x^{\prime}P_{Z}x\right)^{-1}x^{\prime}P_{Z}y\\
 & = & \frac{\sum_{s=1}^{S}n_{s}\overline{x}_{s}\overline{y}_{s}}{\sum_{s=1}^{S}n_{s}\overline{x}_{s}^{2}}\\
 & = & \frac{\sum_{s=1}^{S}n_{s}\overline{x}_{s}^{2}\left(\overline{y}_{s}/\overline{x}_{s}\right)}{\sum_{s=1}^{S}n_{s}\overline{x}_{s}^{2}}=\sum_{s=1}^{S}w_{2sls,s}\widehat{\beta}_{s}\text{,}
\end{eqnarray*}
the standard result that $\widehat{\beta}_{2sls}$ is a linear combination
of the instrument specific IV estimators, (see e.g.\ \citealp{Windmeijer2019}).
The weights are given by
\begin{equation}
w_{2sls,s}=\frac{n_{s}\overline{x}_{s}^{2}}{\sum_{l=1}^{S}n_{l}\overline{x}_{l}^{2}}\geq0\label{ws}
\end{equation}
and hence the 2SLS estimator is here a weighted average of the group
specific estimators.

For the group specific estimates, the first-stage F-statistics are
equal to the Wald statistics for testing the null hypotheses $H_{0}:\pi_{s}=0$,
and are given by
\begin{equation}
F_{\pi_{s}}=\frac{\widehat{\pi}_{s}^{2}}{V\widehat{a}r\left(\widehat{\pi}_{s}\right)}=\frac{n_{s}\overline{x}_{s}^{2}}{\widehat{\sigma}_{v,s}^{2}}\label{Fps}
\end{equation}
for $s=1,...,S$. For each group specific IV estimator $\widehat{\beta}_{s}$
the standard weak instruments results of \citet{StaigerStock1997}
and \citet{StockYogo2005} apply. As these are just-identified models,
we can relate the values of the F-statistics to Wald-test size distortions.

The robust first-stage F-statistic for testing $H_{0}:\pi=0$ is given
by
\begin{eqnarray*}
F_{r} & = & \frac{1}{S}\widehat{\pi}^{\prime}\left(V\widehat{a}r_{r}\left(\widehat{\pi}\right)\right)^{-1}\widehat{\pi}\\
 & = & \frac{1}{S}\sum_{s=1}^{S}\frac{n_{s}\overline{x}_{s}^{2}}{\widehat{\sigma}_{v,s}^{2}}=\frac{1}{S}\sum_{s=1}^{S}F_{\pi_{s}}.
\end{eqnarray*}
It is therefore clear, that if $F_{r}$ is large, then at least one
of the $F_{\pi_{s}}$ is large.

The non-robust F-statistic is given by
\begin{eqnarray*}
F & = & \frac{1}{S}\widehat{\pi}^{\prime}\left(V\widehat{a}r\left(\widehat{\pi}\right)\right)^{-1}\widehat{\pi}\\
 & = & \frac{1}{S}\frac{\sum_{s=1}^{S}n_{s}\overline{x}_{s}^{2}}{\left(\sum_{s=1}^{S}\frac{n_{s}}{n}\widehat{\sigma}_{v,s}^{2}\right)}=\frac{1}{S}\sum_{s=1}^{S}\frac{\widehat{\sigma}_{v,s}^{2}}{\left(\sum_{l=1}^{S}\frac{n_{l}}{n}\widehat{\sigma}_{v,l}^{2}\right)}F_{\pi_{s}}.
\end{eqnarray*}
From (\ref{ws}) and (\ref{Fps}) it follows that the weights for
the 2SLS estimator are related to the individual F-statistics as follows
\[
w_{2sls,s}=\frac{n_{s}\overline{x}_{s}^{2}}{\sum_{l=1}^{S}n_{l}\overline{x}_{l}^{2}}=\frac{\widehat{\sigma}_{v,s}^{2}F_{\pi_{s}}}{\sum_{l=1}^{S}\widehat{\sigma}_{v,l}^{2}F_{\pi_{l}}}.
\]
Under first-stage homoskedasticity, $\sigma_{v,s}^{2}=\sigma_{v,l}^{2}$,
for $s,l=1,\ldots,S$. Then $\widehat{\sigma}_{v,s}^{2}\approx\widehat{\sigma}_{v,l}^{2}$
for all $s,l$, and hence $F\approx\frac{1}{S}\sum_{s=1}^{S}F_{\pi_{s}}$.
Then the weights are given by $w_{2sls,s}\approx\frac{F_{\pi_{s}}}{\sum_{l=1}^{S}F_{\pi_{l}}}\approx\frac{F_{\pi_{s}}}{SF}$,
so we see that then the groups with the larger individual F-statistics
get the larger weights in the 2SLS estimator under homoskedasticity.

This is not necessarily the case under heteroskedasticity. For two
groups with equal value of the F-statistic, the group with the larger
variance gets the larger weight, and indeed, a large variance weakly
identified group could dominate the 2SLS estimator. As shown in the
Monte Carlo exercises below, this is exactly what happens in the design
of \citet{IAndrewsREStat2018}. The robust F-statistic is large because
one of the groups has a large value of the individual F-statistic.
However, this group has a very small variance $\sigma_{v,s}^{2}$
and hence gets a small weight in the 2SLS estimator, resulting in
a poor performance of the estimator in terms of (relative) bias and
Wald-test size.

\section{An Alternative GMM Estimator}

\label{Sec:AltGMM}

Clearly, one would like to use an estimator that gives larger weights
to more strongly identified groups, independent of the value of $\sigma_{v,s}^{2}$,
mimicking the weights of the 2SLS estimator under homoskedasticity
of the first-stage errors. This is achieved by the following GMM estimator,
denoted GMMf, with the extension f for first stage,
\begin{eqnarray*}
\widehat{\beta}_{gmmf} & = & \left(x^{\prime}Z\widehat{\Omega}_{v}^{-1}Z^{\prime}x\right)^{-1}x^{\prime}Z\widehat{\Omega}_{v}^{-1}Z^{\prime}y\\
 & = & \left(\widehat{\pi}^{\prime}Z^{\prime}Z\widehat{\Omega}_{v}^{-1}Z^{\prime}Z\widehat{\pi}\right)^{-1}\widehat{\pi}^{\prime}Z^{\prime}Z\widehat{\Omega}_{v}^{-1}Z^{\prime}y,
\end{eqnarray*}
with $\widehat{\Omega}_{v}=$ $\sum_{i=1}^{n}\widehat{v}_{i}^{2}z_{i}z_{i}^{\prime}$
as defined in (\ref{omev}). This looks like the usual GMM estimator,
but instead of the structural residuals $\widehat{u}$, the first-stage
residuals $\widehat{v}$ are used in the weight matrix. It clearly
links directly to the robust F-statistic, as the denominator is equal
to $SF_{r}$.

It follows that
\begin{eqnarray}
\widehat{\beta}_{gmmf} & = & \frac{\sum_{s=1}^{S}n_{s}\overline{x}_{s}\overline{y}_{s}/\widehat{\sigma}_{v,s}^{2}}{\sum_{s=1}^{S}n_{s}\overline{x}_{s}^{2}/\widehat{\sigma}_{v,s}^{2}}=\frac{\sum_{s=1}^{S}\left(n_{s}\overline{x}_{s}^{2}/\widehat{\sigma}_{v,s}^{2}\right)\widehat{\beta}_{s}}{\sum_{s=1}^{S}n_{s}\overline{x}_{s}^{2}/\widehat{\sigma}_{v,s}^{2}}\label{GMMv}\\
 & = & \sum_{s=1}^{S}w_{gmmf,s}\widehat{\beta}_{s},\nonumber
\end{eqnarray}
with
\[
w_{gmmf,s}=\frac{F_{\pi_{s}}}{\sum_{l=1}^{S}F_{\pi_{l}}}=\frac{F_{\pi_{s}}}{SF_{r}},
\]
and hence the groups with the larger F-statistics get the larger weights,
independent of the values of $\sigma_{v,s}^{2}$, mimicking the 2SLS
weights under homoskedasticity of the first-stage errors.

\section{Weak Instrument Asymptotics}

\label{Sec:WIA}

We can formalize the results obtained above further using weak instruments
asymptotics (WIA). For each group $s=1,...,S$ define
\[
\pi_{s}=\frac{c_{s}}{\sqrt{n_{s}}}.
\]
The limit for $n_{s}\rightarrow\infty$, $s=1,\ldots,S$, of the group
specific concentration parameters are then given by
\begin{equation}
\mu_{s}^{2}=\frac{c_{s}^{2}}{\sigma_{v,s}^{2}}.\label{mus2}
\end{equation}
Then
\[
\widehat{\pi}_{s}=\overline{x}_{s}=\frac{1}{n_{s}}\sum_{j=1}^{n_{s}}\left(\frac{c_{s}}{\sqrt{n_{s}}}+v_{js}\right)=\frac{c_{s}}{\sqrt{n_{s}}}+\overline{v}_{s},
\]
and
\begin{eqnarray*}
n_{s}\overline{x}_{s}^{2} & = & n_{s}\left(\frac{c_{s}}{\sqrt{n_{s}}}+\overline{v}_{s}\right)^{2}\\
 & = & \left(c_{s}^{2}+2c_{s}\sqrt{n_{s}}\overline{v}_{s}+\left(\sqrt{n_{s}}\overline{v}_{s}\right)^{2}\right)\\
 &  & \overset{d}{\rightarrow}\left(c_{s}+\sigma_{v,s}\mathcal{T}_{s}\right)^{2}=\sigma_{v,s}^{2}\left(\mu_{s}+\mathcal{T}_{s}\right)^{2}
\end{eqnarray*}
where $\mu_{s}=c_{s}/\sigma_{v,s}$ and $\mathcal{T}_{s}\sim N\left(0,1\right)$.
We get the standard WIA result that
\[
F_{\pi_{s}}=\frac{n_{s}\overline{x}_{s}^{2}}{\widehat{\sigma}_{v,s}^{2}}\overset{d}{\rightarrow}\left(\mu_{s}+\mathcal{T}_{s}\right)^{2}\sim\chi_{1,\mu_{s}^{2}}^{2},
\]
where $\chi_{1,\mu_{s}^{2}}^{2}$ is the non-central chi-squared distribution
with 1 degree of freedom and non-centrality parameter $\mu_{s}^{2}$.

From (\ref{ws}) it then follows that
\[
w_{2sls,s}\overset{d}{\rightarrow}\frac{\sigma_{v,s}^{2}\left(\mu_{s}+\mathcal{T}_{s}\right)^{2}}{\sum_{l=1}^{S}\sigma_{v,l}^{2}\left(\mu_{l}+\mathcal{T}_{l}\right)^{2}},
\]
with the $\mathcal{T}_{l}$ independent $N\left(0,1\right)$ variables,
for $l=1,...,S$.

For the weights $w_{gmmf,s}$,
\[
w_{gmmf,s}\overset{d}{\rightarrow}\frac{\left(\mu_{s}+\mathcal{T}_{s}\right)^{2}}{\sum_{l=1}^{S}\left(\mu_{l}+\mathcal{T}_{l}\right)^{2}}.
\]

Consider for illustration the case where there are two groups. Table
\ref{tab:WIAw} presents some results for the average values of $w_{2sls,1}$
and $w_{gmmf,1}$ after randomly drawing 100,000 values of $\mathcal{T}_{1}$
and $\mathcal{T}_{2}$. In the first row, there is homoskedasticity,
$\sigma_{v,1}^{2}=\sigma_{v,2}^{2}=5$, and both groups have equal
concentration parameters, $\mu_{1}^{2}=\mu_{2}^{2}=5.76$, which is
the value of the concentration parameter for the group-specific Wald
tests to have a maximal rejection frequency of 10\% at the 5\% level.
Then $E\left(w_{2sls,1}\right)=E\left(w_{gmmf,1}\right)=0.5$ and
both estimators will give on average equal weight to the group specific
estimators.

\begin{table}[H]

\caption{\label{tab:WIAw} WIA weights for 2SLS and GMMf}

\begin{centering}
\begin{tabular}{ccccccc}
\hline
$\sigma_{v,1}^{2}$ & $\sigma_{v,2}^{2}$ & $\mu_{1}^{2}$ & $\mu_{2}^{2}$ &  & $w_{2sls,1}$ & $w_{gmmf,1}$\tabularnewline
\hline
5 & 5 & 5.76 & 5.76 &  & 0.50 & 0.50\tabularnewline
5 & 0.1 & 5.76 & 5.76 &  & 0.95 & 0.50\tabularnewline
5 & 0.1 & 1.96 & 5.76 &  & 0.84 & 0.32\tabularnewline
\hline
\end{tabular}
\par\end{centering}
\centering{}{\small{}Notes: Average weights from $100,000$ draws
of $\mathcal{T}_{1}$ and $\mathcal{T}_{2}$.}{\small\par}
\end{table}

The second row considers the case where there is a large difference
in the variances, $\sigma_{v,1}^{2}=5$, and $\sigma_{v,2}^{2}=0.1$,
but $\mu_{1}^{2}=\mu_{2}^{2}=5.76$ as before. We find for this case
that $E\left(w_{2sls,1}\right)=0.95$, i.e.\ almost all weight will
on average be given to the high variance group 1. The expected weight
for the GMMf estimator is in this case not affected by the relative
values of the $\sigma_{v,s}^{2}$ and remains at $E\left(w_{gmmf,1}\right)=0.5$.
If we subsequently reduce the value of $c_{1}$ such that $\mu_{1}^{2}=1.96$,
then $E\left(w_{2sls,1}\right)=0.84$, i.e.\ the 2SLS estimator will
give more weight to $\widehat{\beta}_{1}$, the estimator in the group
with the smaller concentration parameter, but larger variance. In
contrast, $E\left(w_{gmmf,1}\right)=0.32$ for this case, giving less
weight to the less informative group.

\section{Variance of $u$}

\label{Sec:varu}

So far, focus has been on first-stage heteroskedasticity, with the
robust GMMf estimator exploiting the first-stage information by assigning
larger weights to the groups with larger group specific concentration
parameters independent of the values of $\sigma_{v,s}^{2}$. Consider
next the infeasible robust GMM group IV estimator, given by
\begin{eqnarray*}
\widehat{\beta}_{gmm} & = & \frac{\sum_{s=1}^{S}n_{s}\overline{x}_{s}\overline{y}_{s}/\sigma_{u,s}^{2}}{\sum_{s=1}^{S}n_{s}\overline{x}_{s}^{2}/\sigma_{u,s}^{2}}=\frac{\sum_{s=1}^{S}\left(n_{s}\overline{x}_{s}^{2}/\sigma_{u,s}^{2}\right)\widehat{\beta}_{s}}{\sum_{s=1}^{S}n_{s}\overline{x}_{s}^{2}/\sigma_{u,s}^{2}}\\
 & = & \sum_{s=1}^{S}w_{gmm,s}\widehat{\beta}_{s}.
\end{eqnarray*}
Whereas $\widehat{\beta}_{gmm}$ is the best, normal, consistent and
efficient estimator under standard asymptotics, from the analysis
above it is clear that the weights may not be optimal under WIA. We
have under WIA that
\[
\frac{n_{s}\overline{x}_{s}^{2}}{\sigma_{u,s}^{2}}\overset{d}{\rightarrow}\frac{\sigma_{v,s}^{2}}{\sigma_{u,s}^{2}}\left(\mu_{s}+\mathcal{T}_{s}\right)^{2}
\]
and so
\[
w_{gmm,s}\overset{d}{\rightarrow}\frac{\frac{\sigma_{v,s}^{2}}{\sigma_{u,s}^{2}}\left(\mu_{s}+\mathcal{T}_{s}\right)^{2}}{\sum_{l=1}^{S}\frac{\sigma_{v,l}^{2}}{\sigma_{u,l}^{2}}\left(\mu_{s}+\mathcal{T}_{s}\right)^{2}}.
\]
Clearly, if $u$ is homoskedastic, $\sigma_{u,s}^{2}=\sigma_{u,l}^{2}$
for $\forall$$s,l$, then the infeasible GMM estimator has the same
WIA limiting distribution as the 2SLS estimator and suffers from the
same problems as described above for 2SLS. If $\sigma_{u,s}^{2}=\kappa\sigma_{v,s}^{2}$
for all $s$ then $\widehat{\beta}_{gmm}$ behaves like the GMMf estimator,
the latter in that case also the efficient estimator under standard
asymptotics. For other cases the behaviour of $\widehat{\beta}_{gmm}$
depends on whether $\sigma_{v,s}^{2}/\sigma_{u,s}^{2}$ assigns relatively
larger or smaller weights to the more informative groups.

An alternative is to weight by $\sigma_{u,s}^{2}\sigma_{v,s}^{2}$,
such that
\begin{eqnarray*}
\widehat{\beta}_{gmmuf} & = & \sum_{s=1}^{S}w_{gmmuf,s}\widehat{\beta}_{s},\\
w_{gmmuf,s} & = & \frac{n_{s}\overline{x}_{s}^{2}/\left(\sigma_{u,s}^{2}\sigma_{v,s}^{2}\right)}{\sum_{l=1}^{S}n_{l}\overline{x}_{l}^{2}/\left(\sigma_{u,l}^{2}\sigma_{v,l}^{2}\right)}\overset{d}{\rightarrow}\frac{\frac{1}{\sigma_{u,s}^{2}}\left(\mu_{s}+\mathcal{T}_{s}\right)^{2}}{\sum_{l=1}^{S}\frac{1}{\sigma_{u,l}^{2}}\left(\mu_{l}+\mathcal{T}_{l}\right)^{2}}.
\end{eqnarray*}
The resulting weights are then as for the standard GMM estimator under
first-stage homoskedasticity. This would clearly improve efficiency
if $\sigma_{u,s}^{2}$ is relatively small for the more informative
groups, but can assign again less weight to more informative groups
if their values of $\sigma_{u,s}^{2}$ are relatively large.

\section{Some Monte Carlo Results}

\label{Sec:MC}

We consider here the heteroskedastic design of \citet{IAndrewsREStat2018}
with $S=10$ groups, $\beta=0$ and moderate endogeneity. Table 9
in the Supplementary Appendix C.3 of \citet{IAndrewsREStat2018} presents
the values of the conditional group specific variance matrices $\Sigma_{s}$
as defined in (\ref{eq:Sigmas}) and the first-stage parameters, denoted
$\pi_{0s}$, for $s=1,\ldots,10$. Results for the high endogeneity
case are given in Appendix \ref{subsec:HighEnd}. We multiply the
first-stage parameters $\pi_{0}$ by $0.04$, such that the value
of the robust $F_{r}$ is just over $80$ on average for $10,000$
replications and sample size $n=10,000$. The group sizes are equal
in expectation with $P\left(z_{i}=e_{s}\right)=0.1$ for all $s$.
The first two rows of Table \ref{tab:grinf} present the values of
$\pi_{s}$ and $\sigma_{v,s}^{2}$ for $s=1,\ldots,10$.

Table \ref{tab:resmod} presents the estimation results. The non-robust
F-statistic is small, $F=1.41$ and the effective F-statistic of \citet{OleaPflueger2013},
denoted $F_{eff}$, is equal to the non-robust F in this grouped-data
IV design. Although the robust F-statistic is large, $F_{r}=80.23$,
the 2SLS estimator $\widehat{\beta}_{2sls}$ is poorly behaved. Its
relative bias equal to $0.699$ and the Wald test rejection frequency
for $H_{0}:\beta=0$ is equal to $0.534$ at the 5\% level. In contrast,
the GMMf estimator is unbiased and its Wald-test rejection frequency
equal to $0.049$ at the 5\% level.

\begin{table}[H]

\caption{\label{tab:resmod} Estimation results for $S=10$, moderate endogeneity}

\begin{centering}
\begin{tabular}{cccccccc}
\hline
$F$ & $F_{eff}$ & $F_{r}$ & $\widehat{\beta}_{OLS}$ & $\widehat{\beta}_{2sls}$ & $\widehat{\beta}_{gmmf}$ & $W_{2sls}$ & $W_{gmmf}$\tabularnewline
\hline
\multicolumn{1}{r}{1.411} & \multicolumn{1}{r}{1.411} & \multicolumn{1}{r}{80.23} & \multicolumn{1}{r}{-0.608} & \multicolumn{1}{r}{-0.424} & \multicolumn{1}{r}{-0.001} & \multicolumn{1}{r}{0.534} & \multicolumn{1}{r}{0.049}\tabularnewline
\multicolumn{1}{r}{} & \multicolumn{1}{r}{} & \multicolumn{1}{r}{} & \multicolumn{1}{r}{{\small{}(0.011)}} & \multicolumn{1}{r}{{\small{}(0.257)}} & \multicolumn{1}{r}{{\small{}(0.563)}} & \multicolumn{1}{r}{} & \multicolumn{1}{r}{}\tabularnewline
\hline
\end{tabular}
\par\end{centering}
\centering{}{\small{}Notes: means and (st.dev.) of $10,000$ replications.
Rej.freq. of robust Wald tests at 5\% level. }{\small\par}
\end{table}

The details as given in Table \ref{tab:grinf} below make clear what
is happening. It reports the fixed values of $\pi_{s}$, $\sigma_{v,s}^{2}$,
$\mu_{n,s}^{2}=1000\pi_{s}^{2}/\sigma_{v,s}^{2}$ and the mean values
of $F_{\pi_{s}}$, $w_{2sls,s}$ and $w_{gmmf,s}=F_{\pi_{s}}/\sum_{l=1}^{S}F_{\pi_{l}}$.
Identification in the first group is strong, with an average value
of $F_{\pi_{1}}=789.5$. Identification in all other 9 groups is very
weak, with the largest average value for $F_{\pi_{5}}=2.23$. But
the variance in group 1 is very small, and some of the variances in
the other groups are quite large. This leads to the low average value
of $w_{2sls,1}=0.127$, showing that the 2SLS estimator does not utilize
the identification strength of the first group, with larger weight
given to higher variance, but lower concentration-parameter groups.

\begin{table}[H]

\caption{\label{tab:grinf}Group information and estimator weights}

\begin{centering}
\begin{tabular}{lcccccccccc}
\hline
$s$ & 1 & 2 & 3 & 4 & 5 & 6 & 7 & 8 & 9 & 10\tabularnewline
\hline
$\pi_{s}$ & \multicolumn{1}{r}{0.058} & \multicolumn{1}{r}{-0.023} & \multicolumn{1}{r}{0.049} & \multicolumn{1}{r}{0.015} & \multicolumn{1}{r}{0.022} & \multicolumn{1}{r}{0.008} & \multicolumn{1}{r}{-0.017} & \multicolumn{1}{r}{0.011} & \multicolumn{1}{r}{-0.036} & \multicolumn{1}{r}{-0.040}\tabularnewline
$\sigma_{v,s}^{2}$ & \multicolumn{1}{r}{0.004} & \multicolumn{1}{r}{2.789} & \multicolumn{1}{r}{4.264} & \multicolumn{1}{r}{0.779} & \multicolumn{1}{r}{0.395} & \multicolumn{1}{r}{7.026} & \multicolumn{1}{r}{1.226} & \multicolumn{1}{r}{0.308} & \multicolumn{1}{r}{1.709} & \multicolumn{1}{r}{6.099}\tabularnewline
 & \multicolumn{1}{r}{} & \multicolumn{1}{r}{} & \multicolumn{1}{r}{} & \multicolumn{1}{r}{} & \multicolumn{1}{r}{} & \multicolumn{1}{r}{} & \multicolumn{1}{r}{} & \multicolumn{1}{r}{} & \multicolumn{1}{r}{} & \multicolumn{1}{r}{}\tabularnewline
$\mu_{n,s}^{2}$ & \multicolumn{1}{r}{785.7} & \multicolumn{1}{r}{0.184} & \multicolumn{1}{r}{0.556} & \multicolumn{1}{r}{0.284} & \multicolumn{1}{r}{1.190} & \multicolumn{1}{r}{0.009} & \multicolumn{1}{r}{0.236} & \multicolumn{1}{r}{0.387} & \multicolumn{1}{r}{0.770} & \multicolumn{1}{r}{0.266}\tabularnewline
$F_{\pi_{s}}$ & \multicolumn{1}{r}{789.5} & \multicolumn{1}{r}{1.170} & \multicolumn{1}{r}{1.564} & \multicolumn{1}{r}{1.279} & \multicolumn{1}{r}{2.225} & \multicolumn{1}{r}{0.997} & \multicolumn{1}{r}{1.203} & \multicolumn{1}{r}{1.372} & \multicolumn{1}{r}{1.798} & \multicolumn{1}{r}{1.246}\tabularnewline
 &  &  &  &  &  &  &  &  &  & \tabularnewline
$w_{2sls,s}$ & \multicolumn{1}{r}{0.126} & \multicolumn{1}{r}{0.098} & \multicolumn{1}{r}{0.178} & \multicolumn{1}{r}{0.035} & \multicolumn{1}{r}{0.031} & \multicolumn{1}{r}{0.180} & \multicolumn{1}{r}{0.049} & \multicolumn{1}{r}{0.015} & \multicolumn{1}{r}{0.096} & \multicolumn{1}{r}{0.192}\tabularnewline
$w_{gmmf,s}$ & \multicolumn{1}{r}{0.984} & \multicolumn{1}{r}{0.002} & \multicolumn{1}{r}{0.002} & \multicolumn{1}{r}{0.002} & \multicolumn{1}{r}{0.003} & \multicolumn{1}{r}{0.001} & \multicolumn{1}{r}{0.002} & \multicolumn{1}{r}{0.002} & \multicolumn{1}{r}{0.002} & \multicolumn{1}{r}{0.002}\tabularnewline
\hline
\end{tabular}
\par\end{centering}
\begin{centering}
{\small{}Notes: }$\mu_{n,s}^{2}=1000\pi_{s}^{2}/\sigma_{v,s}^{2}$.
\par\end{centering}
\end{table}

Table \ref{tab:grinf} further shows that for the GMMf estimator almost
all weight is given to the first group, with the average of $w_{gmmf,1}$
equal to $0.984$, resulting in the good behaviour of the GMMf estimator
in terms of bias and Wald test size. In this case the standard deviation
of the GMMf estimator is quite large relative to that of the 2SLS
estimator. This is driven by the value of $\sigma_{u,1}^{2}$, which
in this design is equal to $1.10$, much larger than $\sigma_{v,1}^{2}$.
Reducing the value of $\sigma_{u,1}^{2}$ (and the value for $\sigma_{uv,1}$
accordingly to keep the same correlation structure within group 1),
will reduce the standard deviation of the GMMf estimator.

\begin{figure}
\begin{centering}
\includegraphics[scale=0.55]{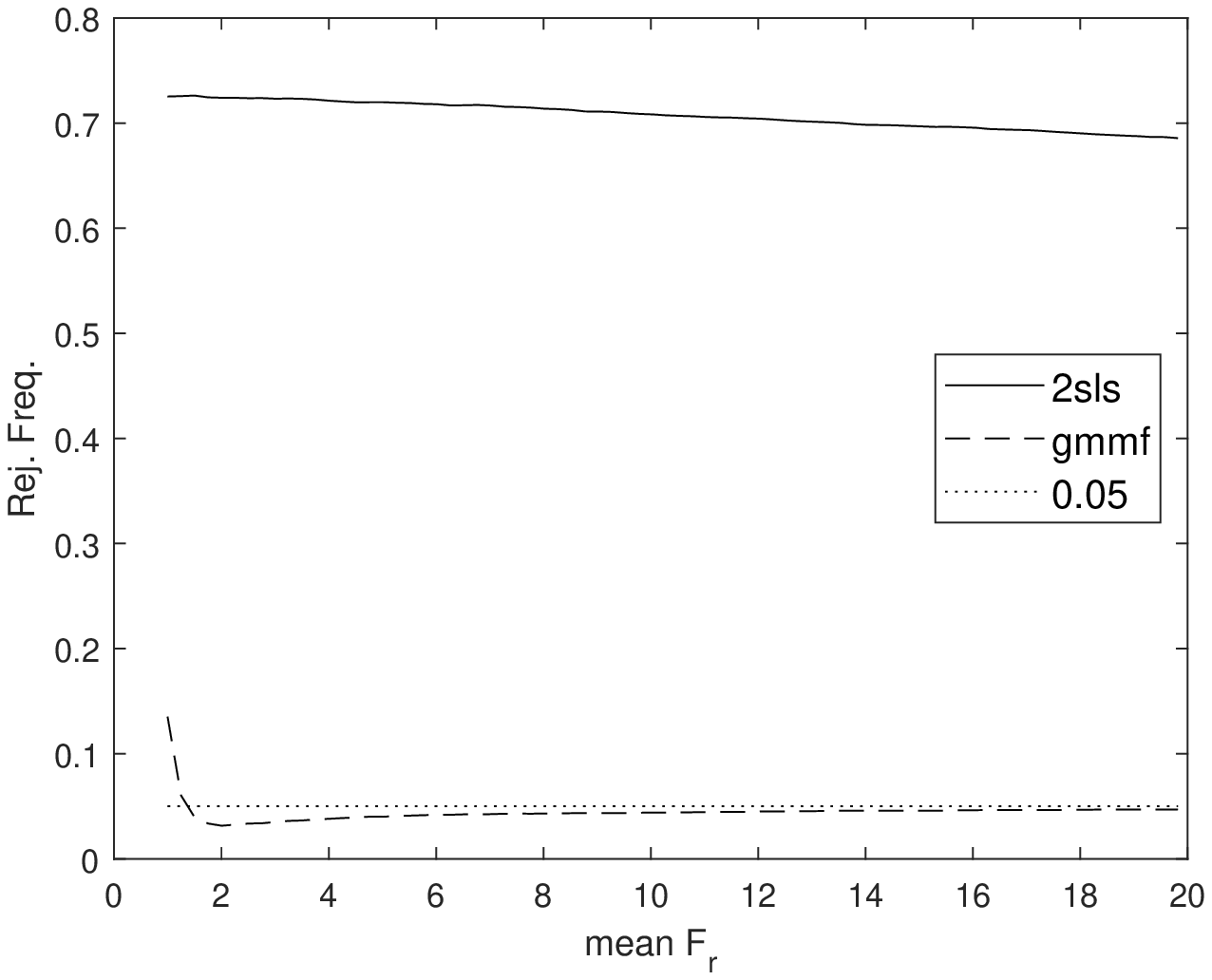}\includegraphics[scale=0.55]{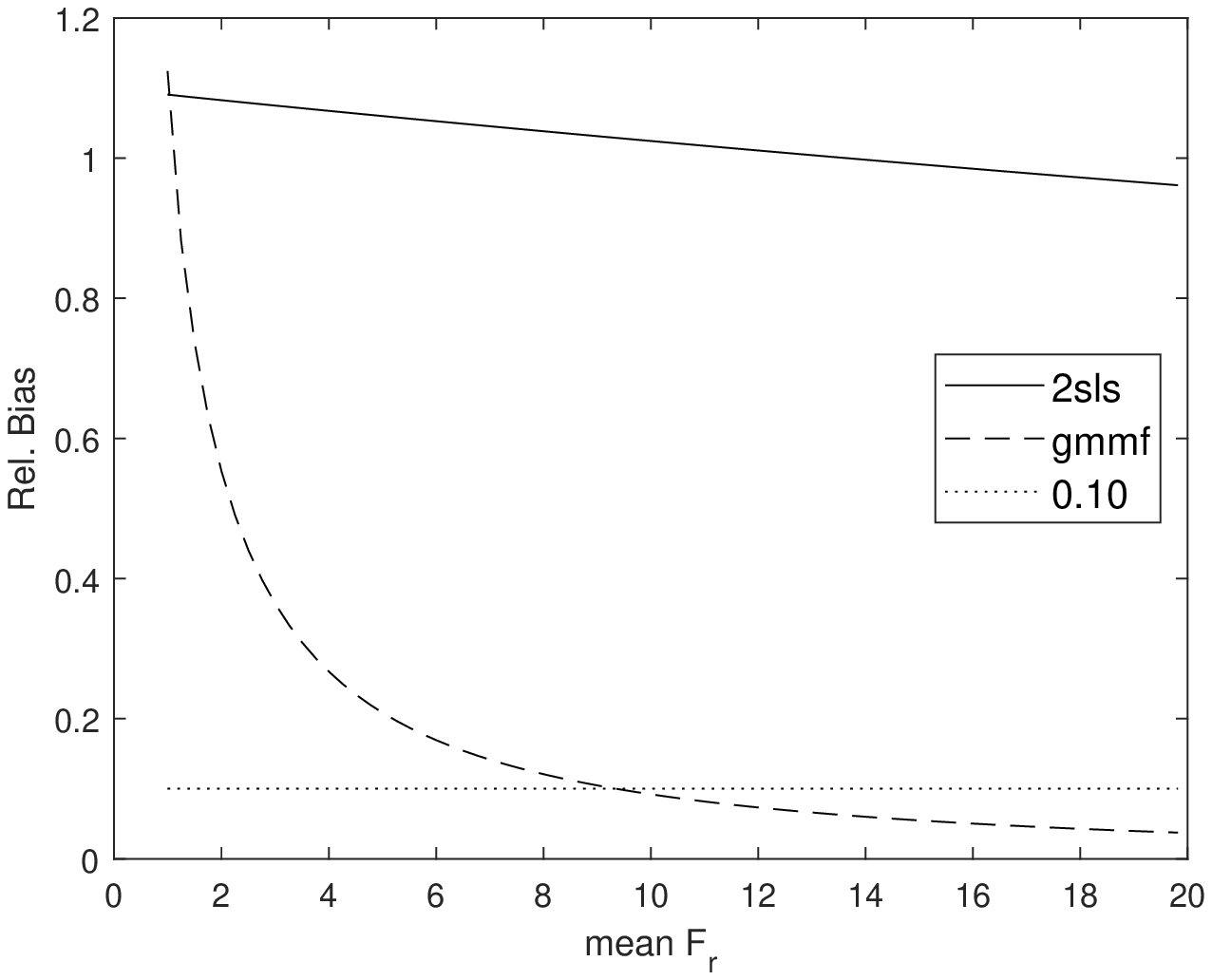}
\par\end{centering}
\caption{\label{fig:rejfrelb} Rejection frequencies of robust Wald tests and
relative biases of 2SLS and GMMf estimators}

\end{figure}

Figure \ref{fig:rejfrelb} displays the rejection frequencies of the
robust Wald tests for testing $H_{0}:\beta=0$ for varying values
of the robust F-statistic $F_{r}$ for the 2SLS and GMMf estimators.
Different values of $F_{r}$ are obtained by different values of $d$
when setting the first-stage parameters $\pi=d\pi_{0}$. It is clear
that the Wald test based on the GMMf estimator is much better behaved
in terms of size than the test based on the 2SLS estimator, with hardly
any size distortion for mean values of $F_{r}$ larger than 5. The
right panel of Figure \ref{fig:rejfrelb} shows that the bias of the
GMMf estimator, relative to that of the OLS estimator, is also substantially
smaller than that of the 2SLS estimator, with the relative bias smaller
than $0.10$ for mean values of $F_{r}$ larger than 9.

\section{Dynamic Panel Data Model}

\label{Sec:Panel}

Next, consider the dynamic AR(1) panel data specification
\begin{equation}
y_{it}=\gamma y_{i,t-1}+\eta_{i}+u_{it},\label{ar1pan}
\end{equation}
for $i=1,...,n$, and $t=2,..,T$, and for $\left|\gamma\right|<1$.
Let $y_{it}^{\ast}$ and $y_{i,t-1}^{\ast}$ be the forward orthogonal
deviations transformed variables, see \citet{ArellanoBover1995},
and $y_{i}^{\ast}$ and $y_{i,-1}^{\ast}$ the associated $\left(T-2\right)$
vectors. Let the $\left(T-2\right)\times\left(T-1\right)\left(T-2\right)/2$
matrix of instruments $Z_{i}$ be defined as
\[
Z_{i}=\left[\begin{array}{cccccccc}
y_{i1} & 0 & 0 & 0 & 0 & 0 & 0 & 0\\
0 & y_{i1} & y_{i2} & 0 & 0 & 0 & 0 & 0\\
 &  &  & \ddots\\
0 & 0 & 0 & 0 & y_{i1} & y_{i2} & \ldots & y_{i,T-2}
\end{array}\right].
\]
Further, let the $n\left(T-2\right)$ vectors $y^{\ast}=\left(y_{1}^{\ast\prime},y_{2}^{\ast\prime},...,y_{n}^{\ast\prime}\right)^{\prime}$
and $y_{-1}^{\ast}=\left(y_{1,-1}^{\ast\prime},y_{2,-1}^{\ast\prime},...,y_{n,-1}^{\ast\prime}\right)^{\prime}$,
and the $n\left(T-2\right)\times\left(T-1\right)\left(T-2\right)/2$
instrument matrix $Z=\left[Z_{1}^{\prime},Z_{2}^{\prime},...,Z_{n}^{\prime}\right]^{\prime}$.
The 2SLS estimator is then given by
\[
\widehat{\gamma}_{2sls}=\left(y_{-1}^{\ast\prime}Z\left(Z^{\prime}Z\right)^{-1}Z^{\prime}y_{-1}^{\ast}\right)^{-1}y_{-1}^{\ast\prime}Z\left(Z^{\prime}Z\right)^{-1}Z^{\prime}y^{\ast}
\]
and is consistent and asymptotic normal under standard asymptotics,
regularity assumptions and the assumption of no serial correlation
in $u_{it}$, $E\left(u_{it}u_{is}\right)=0$ for $t\neq s$. It is
efficient under conditional homoskedasticity, $E\left(u_{i}u_{i}^{\prime}|Z_{i}\right)=\sigma_{u}^{2}I_{T-1}$.

The first stage for the 2SLS estimator is here given by
\[
y_{i,-1}^{\ast}=Z_{i}^{\prime}\pi+v_{i}^{\ast}.
\]
The difference here compared to the grouped data IV example is that
the $v_{it}^{\ast}$ are not drawn separately, but the processes are
driven by the $u_{it}$ only. Let $\widehat{\pi}$ be the OLS estimator
of $\pi$, $\widehat{v}_{i}^{\ast}=y_{i,-1}^{\ast}-Z_{i}^{\prime}\widehat{\pi}$
and
\[
\widehat{\Omega}_{v}^{\ast}=\sum_{i=1}^{n}Z_{i}^{\prime}\widehat{v}_{i}^{\ast}\widehat{v}_{i}^{\ast\prime}Z_{i}.
\]
Then the robust F-statistic for $H_{0}:\pi=0$ is given by
\[
F_{r}=y_{-1}^{\ast\prime}Z\left(\widehat{\Omega}_{v}^{\ast}\right)^{-1}Z^{\prime}y_{-1}^{\ast}/k_{z},
\]
where $k_{z}=\left(T-1\right)\left(T-2\right)/2$. Accordingly, the
GMMf estimator is here given by
\[
\widehat{\gamma}_{gmmf}=\left(y_{-1}^{\ast\prime}Z\left(\widehat{\Omega}_{v}^{\ast}\right)^{-1}Z^{\prime}y_{-1}^{\ast}\right)^{-1}y_{-1}^{\ast\prime}Z\left(\widehat{\Omega}_{v}^{\ast}\right)^{-1}Z^{\prime}y^{\ast}.
\]

Under conditional heteroskedasticity the standard two-step GMM estimator
is efficient under standard asymptotics. Denote $\widehat{u}_{i}^{\ast}=y_{i}^{\ast}-\widehat{\gamma}_{2sls}y_{i,-1}^{\ast}$,
and let
\[
\widehat{\Omega}_{u}^{\ast}=\sum_{i=1}^{n}Z_{i}^{\prime}\widehat{u}_{i}^{\ast}\widehat{u}_{i}^{\ast\prime}Z_{i}.
\]
The two-step GMM estimator is then given by
\[
\widehat{\gamma}_{gmm2}=\left(y_{-1}^{\ast\prime}Z\left(\widehat{\Omega}_{u}^{\ast}\right)^{-1}Z^{\prime}y_{-1}^{\ast}\right)^{-1}y_{-1}^{\ast\prime}Z\left(\widehat{\Omega}_{u}^{\ast}\right)^{-1}Z^{\prime}y^{\ast}.
\]

Following \citet[p 154]{Arellano2003}, the 2SLS estimator can also
be obtained as a weighted average of cross-sectional 2SLS estimators.
\begin{eqnarray*}
\widehat{\gamma}_{2sls} & = & \left(\sum_{t=2}^{T-1}y_{t,-1}^{\ast\prime}Z_{t}\left(Z_{t}^{\prime}Z_{t}\right)^{-1}Z_{t}^{\prime}y_{t,-1}^{\ast}\right)^{-1}\sum_{t=2}^{T-1}y_{t,-1}^{\ast\prime}Z_{t}\left(Z_{t}^{\prime}Z_{t}\right)^{-1}Z_{t}^{\prime}y_{t}^{\ast}\\
 & = & \sum_{t=2}^{T-1}w_{2sls,t}\widehat{\gamma}_{t},
\end{eqnarray*}
where
\begin{eqnarray*}
\widehat{\gamma}_{t} & = & \left(y_{t,-1}^{\ast\prime}Z_{t}\left(Z_{t}^{\prime}Z_{t}\right)^{-1}Z_{t}^{\prime}y_{t,-1}^{\ast}\right)^{-1}y_{t,-1}^{\ast\prime}Z_{t}\left(Z_{t}^{\prime}Z_{t}\right)^{-1}Z_{t}^{\prime}y_{t}^{\ast}\\
w_{2sls,t} & = & \frac{\left(y_{t,-1}^{\ast\prime}Z_{t}\left(Z_{t}^{\prime}Z_{t}\right)^{-1}Z_{t}^{\prime}y_{t,-1}^{\ast}\right)}{\sum_{l=2}^{T-1}y_{l,-1}^{\ast\prime}Z_{l}\left(Z_{l}^{\prime}Z_{l}\right)^{-1}Z_{l}^{\prime}y_{l,-1}^{\ast}},
\end{eqnarray*}
with the $n$-vectors $y_{t}^{\ast}=\left(y_{1t}^{\ast},...,y_{nt}^{\ast}\right)^{\prime}$,
$y_{t,-1}^{\ast}=$ $\left(y_{1,t-1}^{\ast},...,y_{n,t-1}^{\ast}\right)^{\prime}$,
and the $n\times\left(t-1\right)$ matrix $Z_{t}=\left[\begin{array}{ccc}
y_{1} & \ldots & y_{t-1}\end{array}\right]$, with $y_{t}=\left(y_{1t},...,y_{nt}\right)^{\prime}$. We therefore
see that we are in a similar setup as the grouped-data IV example,
with the group-specific IV estimators here the cross-section specific
ones. The 2SLS estimator may therefore again give too much weight
to less informative groups if the associated cross-sectional variance
of the first-stage error in the forward orthogonal deviations transformed
model is large.

We illustrate this with the following design. We set $n=200$, $T=5$,
$\gamma=0.9$, and draw $\eta_{i}\sim N\left(0,1\right)$ and $y_{i0}\sim N\left(\frac{\eta_{i}}{1-\gamma},\frac{1}{1-\gamma^{2}}\right)$.
The data are then generated according to model (\ref{ar1pan}) for
$t=1,..,5$. The $u_{it}$ are independently drawn, $u_{it}\sim N\left(0,\sigma_{u,t}^{2}\right)$,
and so are iid at the cross sectional level. Table 4 presents the
estimation results for the design with $\sigma_{u,t}=1$ for $t=1,2,4,5$,
whereas $\sigma_{u,3}=4$.

The OLS estimator of $\gamma$ in the transformed model $y^{\ast}=\gamma y_{-1}^{\ast}+u^{\ast}$
is denoted $\widehat{\gamma}_{ols}$. Whereas the GMMf estimator takes
fully account of the clustering of the first-stage errors, an alternative
estimator denoted $\widehat{\gamma}_{\widehat{\sigma}_{v}^{2}}$ takes
account of the period specific variances only and is defined as
\[
\widehat{\gamma}_{\widehat{\sigma}_{v}^{2}}=\sum_{t=2}^{T-1}w_{\widehat{\sigma}_{v}^{2},t}\widehat{\gamma}_{t},
\]
with
\[
w_{\widehat{\sigma}_{v}^{2},t}=\frac{\left(y_{t,-1}^{\ast\prime}Z_{t}\left(Z_{t}^{\prime}Z_{t}\right)^{-1}Z_{t}^{\prime}y_{t,-1}^{\ast}\right)/\widehat{\sigma}_{v,t}^{2}}{\sum_{l=2}^{T-1}y_{l,-1}^{\ast\prime}Z_{l}\left(Z_{l}^{\prime}Z_{l}\right)^{-1}Z_{l}^{\prime}y_{l,-1}^{\ast}/\widehat{\sigma}_{v,l}^{2}},
\]
where $\widehat{\sigma}_{v,t}^{2}=\widehat{v}_{t}^{\prime}\widehat{v}_{t}/n$,
with $\widehat{v}_{t}=y_{t,-1}^{\ast}-Z_{t}\widehat{\pi}_{t}$.

Estimation results for this design are presented in Table \ref{tab:AR1}.
Due to the time heteroskedasticity, the robust F-statistic has a larger
mean than the non-robust F-statistic, 6.74 vs 1.44. The OLS estimator
is severely downward biased. The 2SLS estimator also has a large downward
bias and its relative bias is 0.70. The two-step GMM estimator is
less biased, but still has a large relative bias of 0.45. The GMMf
and $\widehat{\gamma}_{\widehat{\sigma}_{v}^{2}}$ estimators perform
better in terms of bias, and have relative biases of 0.25 and 0.16
respectively.

\begin{table}[H]
\caption{\label{tab:AR1}Estimation results for AR(1) panel data model}

\begin{centering}
\begin{tabular}{cccccccc}
\hline
$F$ & $F_{r}$ & $\widehat{\gamma}_{ols}$ & $\widehat{\gamma}_{2sls}$ & $\widehat{\gamma}_{gmm2}$ & $\widehat{\gamma}_{gmmf}$ & $\widehat{\gamma}_{\widehat{\sigma}_{v}^{2}}$ & \tabularnewline
\hline
1.440 & 6.741 & 0.371 & 0.527 & 0.662 & 0.768 & 0.815 & \tabularnewline
 &  & {\small{}(0.017)} & {\small{}(0.231)} & {\small{}(0.254)} & {\small{}(0.287)} & {\small{}(0.219)} & \tabularnewline
\hline
\end{tabular}
\par\end{centering}
\centering{}{\small{}Notes: }$\gamma=0.9${\small{}, }$n=200${\small{},
}$T=5${\small{}. Means and (st.dev.) from }$10,000${\small{} MC
replications.}{\small\par}
\end{table}

Table \ref{tab:panelgrinf} displays the time specific information
and paints a similar picture as that given in Table \ref{tab:grinf}
for the grouped-data IV case. The F-statistics for $t=2$ and $t=3$
are small and the first-stage variances are relatively large. The
average F-statistic for $t=4$ is relatively large, but the first-stage
variance is small. The 2SLS estimator gives a relatively large weight
to the uninformative periods, whereas the first-stage variance weighted
estimator gives a large weight to the informative $t=4$ period.

\begin{table}[H]
\caption{\label{tab:panelgrinf} Period specific information and estimator
weights}

\begin{centering}
\begin{tabular}{llccccl}
\hline
\ \ \ \ \ \ \ \ \  & $t$ & \ \ \ \ \ \ \ \ \  & 2 & 3 & 4 & \tabularnewline
\hline
 & $\widehat{\sigma}_{v,t}^{2}$ &  & 5.620 & 9.780 & 0.516 & \tabularnewline
 & $F_{t}$ &  & 1.268 & 1.099 & 11.14 & \tabularnewline
 & $w_{2sls,t}$ &  & 0.126 & 0.393 & 0.480 & \tabularnewline
 & $w_{\widehat{\sigma}_{v}^{2},t}$ &  & 0.035 & 0.062 & 0.903 & \tabularnewline
\hline
\end{tabular}
\par\end{centering}
\centering{}{\small{}Notes: Averages over $10,000$ MC replications}{\small\par}
\end{table}

\begin{figure}[h]
\begin{centering}
\includegraphics[scale=0.65]{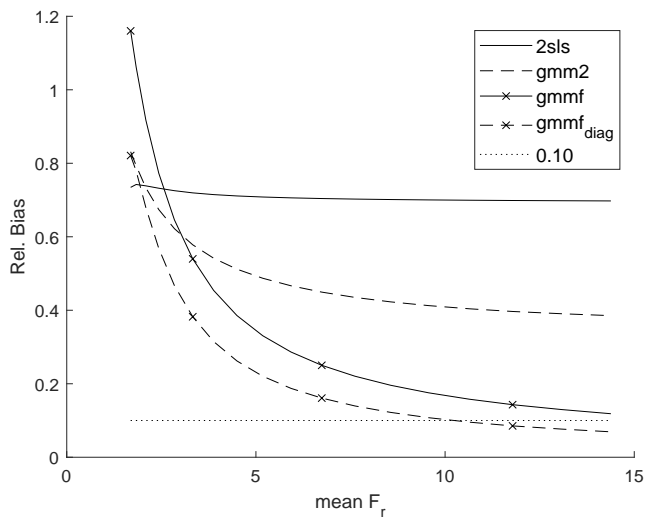}
\par\end{centering}
\caption{\label{fig:Relbias}Relative bias}

\end{figure}

Figure \ref{fig:Relbias} displays the relative bias of the estimators
for varying values of the robust F-statistic $F_{r}$ by varying the
value of $\sigma_{u,3}=\left\{ 1,1.3,1.6...,6.1\right\} $. We see
again that the GMMf and $\widehat{\gamma}_{\widehat{\sigma}_{v}^{2}}$
(GMMf$_{diag}$) estimators utilize the information as conveyed by
the robust F-statistic well. The 2SLS estimator has a large relative
bias for all values of the mean of $F_{r}$. Whereas the two-step
GMM estimator's performance improves with increasing value of $F_{r}$
in this setting, its relative bias remains high at 0.385 at the largest
mean value of $F_{r}$ considered, $14.38$. The relative biases of
the GMMf and $\widehat{\gamma}_{\widehat{\sigma}_{v}^{2}}$ estimators
are respectively 0.119 and 0.069 at that value of the mean of $F_{r}$.

\section{Testing for Weak Instruments}

\label{Sec:wit}

Using the GMMf estimator as a generalization of the 2SLS estimator
to deal with general forms of first-stage heteroskedasticity, we derive
in the Appendix under what conditions the weak-instruments \citet{StockYogo2005}
critical values derived for the non-robust F-test and the properties
of the 2SLS estimator under full homoskedasticity apply to the robust
F-test and the properties of the GMMf estimator. We focus here on
standard cross-sectional heteroskedasticity, but results apply to
cluster and/or serially correlated designs.

Consider again the standard linear model
\begin{eqnarray*}
y_{i} & = & x_{i}\beta+u_{i};\\
x_{i} & = & z_{i}^{\prime}\pi+v_{i},
\end{eqnarray*}
where $z_{i}$ is a $k_{z}$-vector of instruments, and where other
exogenous variables, including the constant have been partialled out.
General conditional heteroskedasticity is specified as
\begin{eqnarray*}
E\left[u_{i}^{2}|z_{i}\right] & = & \sigma_{u}^{2}\left(z_{i}\right);\,\,E\left[v_{i}^{2}|z_{i}\right]=\sigma_{v}^{2}\left(z_{i}\right);\,\,E\left[u_{i}v_{i}|z_{i}\right]=\sigma_{uv}\left(z_{i}\right).
\end{eqnarray*}
Further, let
\[
\Omega_{u}=E\left[\sigma_{u}^{2}\left(z_{i}\right)z_{i}z_{i}^{\prime}\right];\ \ \Omega_{v}=E\left[\sigma_{v}^{2}\left(z_{i}\right)z_{i}z_{i}^{\prime}\right];\ \ \Omega_{uv}=E\left[\sigma_{uv}\left(z_{i}\right)z_{i}z_{i}^{\prime}\right],
\]
and the unconditional variances and covariance
\[
\sigma_{u}^{2}=E_{z}\left[\sigma_{u}^{2}\left(z_{i}\right)\right];\ \ \sigma_{v}^{2}=E_{z}\left[\sigma_{v}^{2}\left(z_{i}\right)\right];\ \ \sigma_{uv}=E_{z}\left[\sigma_{uv}\left(z_{i}\right)\right].
\]

The robust F-statistic and GMMf estimator are given by
\begin{eqnarray*}
F_{r} & = & x^{\prime}Z\widehat{\Omega}_{v}^{-1}Z^{\prime}x/k_{z}\\
\widehat{\beta}_{gmmf} & = & \left(x^{\prime}Z\widehat{\Omega}_{v}^{-1}Z^{\prime}x\right)^{-1}x^{\prime}Z\widehat{\Omega}_{v}^{-1}Z^{\prime}y
\end{eqnarray*}
\citet{StockYogo2005} derived critical values for the non-robust
F-statistic under homoskedasticity for the weak-instruments hypothesis
on the relative bias of the 2SLS estimator, relative to that of the
OLS estimator. In Appendix \ref{Sec:AppWIA} we show that these critical
values apply to the robust F-statistic for relative bias of the GMMf
estimator, relative to that of the OLS estimator if $\Omega_{uv}=\delta\Omega_{v}$
and $\sigma_{uv}=\delta\sigma_{v}^{2}$, where $\delta$ is some arbitrary
constant.

For the Wald test size distortion, we show in Appendix \ref{Sec:AppWIA}
that the \citet{StockYogo2005} critical values apply to the GMMf
based Wald test if $\Omega_{uv}=\delta\Omega_{v}$ and $\Omega_{u}=\kappa\Omega_{v}$,
with $\delta$ and $\kappa$ some arbitrary constants. The condition
$\Omega_{u}=\kappa\Omega_{v}$ implies that the GMMf estimator is
also the efficient estimator under standard asymptotics.

Whilst these conditions imply a limited applicability of the \citet{StockYogo2005}
critical values for the robust F-statistic in relation to the behaviour
of the GMMf estimator, it is a generalization of, and includes, the
homoskedastic case. It also encompasses the illustrative example of
\citet[Section 3.1]{OleaPflueger2013}, where they considered a design
with $E\left[\left(u_{i}~v_{i}\right)^{\prime}\left(u_{i}~v_{i}\right)\right]=\Sigma$
and $E\left[\left(\left(u_{i}~v_{i}\right)^{\prime}\left(u_{i}~v_{i}\right)\right)\otimes z_{i}z_{i}^{\prime}\right]=a^{2}\Sigma\otimes I_{k_{z}}$,
and where the non-robust F-statistic gives an overestimate of the
information content for the 2SLS estimator when $a>1$.

\section{Conclusions}

This paper has shown why large values of the first-stage robust F-statistic
may not translate in good behaviour of the 2SLS estimator. In the
heteroskedastic grouped-data design of \citet{IAndrewsREStat2018},
this is the case because a highly informative group had a relatively
small first-stage variance, and the 2SLS estimator gives more weight
to groups with small concentration parameters but large first-stage
variances. A robust GMM estimator, called GMMf, with the robust weight
matrix estimated using the first-stage residuals, remedies this problem
and gives larger weights to more informative groups. This is independent
of the values of the first-stage variances and is a generalization
of the 2SLS estimator in that it mimics what the 2SLS estimator does
under first-stage homoskedasticity. A large value of the robust F-statistic
indicates that there is first-stage information resulting in a well
behaved GMMf estimator, also confirmed in an AR(1) dynamic panel data
model. We have provided the conditions under which the \citet{StockYogo2005}
weak instruments critical values developed for the non-robust F-statistic
and relative bias and Wald test size distortion of the 2SLS estimator
apply to the robust F-statistic and the behaviour of the GMMf estimator.

\global\long\def\thesection{A}%
\global\long\def\theequation{A.\arabic{equation}}%
\setcounter{table}{0}\setcounter{figure}{0}\setcounter{equation}{0}

\global\long\def\thetable{A\arabic{table}}%
\global\long\def\thefigure{A\arabic{figure}}%

\section*{Appendix}

\subsection{Results for high endogeneity design}

\label{subsec:HighEnd}

Tables \ref{tab:gringhi} and \ref{tab:EstiHi} present estimation
results for the $S=10$, high endogeneity design of \citet[SA, Table 12]{IAndrewsREStat2018}.
As in Section \ref{Sec:MC}, the first-stage parameters have been
multiplied by a factor such that the robust F-statistic has an average
value of just over 80. As shown in Table \ref{tab:gringhi}, the pattern
of group information is similar to that in the moderate endogeneity
case, with one informative group, group $s=10$, with an average value
of $F_{\pi_{10}}=792.2$. However, the variance $\sigma_{v,10}^{2}$
is now so small in relative terms, that the 2SLS weight for group
10 has an average value of only $w_{2sls,10}=0.003$. The GMMf estimator
corrects this, with the average value of $w_{gmmf,10}=0.989$, and
is again much better behaved than the 2SLS estimator both in terms
of (relative) bias and Wald test size, as displayed in Table \ref{tab:EstiHi}.

\begin{table}[H]

\caption{\label{tab:gringhi} Group information and estimator weights, high
endogeneity}

\begin{centering}
\begin{tabular}{lcccccccccc}
\hline
$s$ & 1 & 2 & 3 & 4 & 5 & 6 & 7 & 8 & 9 & 10\tabularnewline
\hline
$100\cdot\pi_{s}$ & \multicolumn{1}{r}{{\small{}-0.021}} & \multicolumn{1}{r}{{\small{}0.095}} & \multicolumn{1}{r}{{\small{}-0.484}} & \multicolumn{1}{r}{{\small{}-0.069}} & \multicolumn{1}{r}{{\small{}0.159}} & \multicolumn{1}{r}{{\small{}-0.028}} & \multicolumn{1}{r}{{\small{}0.101}} & \multicolumn{1}{r}{{\small{}-0.418}} & \multicolumn{1}{r}{{\small{}0.450}} & \multicolumn{1}{r}{{\small{}-0.546}}\tabularnewline
$\sigma_{v,s}^{2}$ & \multicolumn{1}{r}{{\small{}1.600}} & \multicolumn{1}{r}{{\small{}0.478}} & \multicolumn{1}{r}{{\small{}2.975}} & \multicolumn{1}{r}{{\small{}1.142}} & \multicolumn{1}{r}{{\small{}0.174}} & \multicolumn{1}{r}{{\small{}0.145}} & \multicolumn{1}{r}{{\small{}4.658}} & \multicolumn{1}{r}{{\small{}1.963}} & \multicolumn{1}{r}{{\small{}2.990}} & \multicolumn{1}{r}{{\small{}0.38}$\cdot a$}\tabularnewline
 & \multicolumn{1}{r}{} & \multicolumn{1}{r}{} & \multicolumn{1}{r}{} & \multicolumn{1}{r}{} & \multicolumn{1}{r}{} & \multicolumn{1}{r}{} & \multicolumn{1}{r}{} & \multicolumn{1}{r}{} & \multicolumn{1}{r}{} & \multicolumn{1}{r}{}\tabularnewline
$\mu_{n,s}^{2}$ & \multicolumn{1}{r}{{\small{}0.28}$\cdot a$} & \multicolumn{1}{r}{{\small{}0.002}} & \multicolumn{1}{r}{{\small{}0.008}} & \multicolumn{1}{r}{{\small{}4.2}$\cdot a$} & \multicolumn{1}{r}{{\small{}0.015}} & \multicolumn{1}{r}{{\small{}5.6}$\cdot a$} & \multicolumn{1}{r}{{\small{}2.2}$\cdot a$} & \multicolumn{1}{r}{{\small{}0.009}} & \multicolumn{1}{r}{{\small{}0.007}} & \multicolumn{1}{r}{{\small{}789.9}}\tabularnewline
$F_{\pi_{s}}$ & \multicolumn{1}{r}{{\small{}0.998}} & \multicolumn{1}{r}{{\small{}1.017}} & \multicolumn{1}{r}{{\small{}0.979}} & \multicolumn{1}{r}{{\small{}1.010}} & \multicolumn{1}{r}{{\small{}1.034}} & \multicolumn{1}{r}{{\small{}0.984}} & \multicolumn{1}{r}{{\small{}0.977}} & \multicolumn{1}{r}{{\small{}1.031}} & \multicolumn{1}{r}{{\small{}0.997}} & \multicolumn{1}{r}{{\small{}792.2}}\tabularnewline
 &  &  &  &  &  &  &  &  &  & \tabularnewline
$w_{2sls,s}$ & \multicolumn{1}{r}{{\small{}0.111}} & \multicolumn{1}{r}{{\small{}0.040}} & \multicolumn{1}{r}{{\small{}0.177}} & \multicolumn{1}{r}{{\small{}0.085}} & \multicolumn{1}{r}{{\small{}0.016}} & \multicolumn{1}{r}{{\small{}0.013}} & \multicolumn{1}{r}{{\small{}0.242}} & \multicolumn{1}{r}{{\small{}0.134}} & \multicolumn{1}{r}{{\small{}0.181}} & \multicolumn{1}{r}{{\small{}0.003}}\tabularnewline
$w_{gmmf,s}$ & \multicolumn{1}{r}{{\small{}0.001}} & \multicolumn{1}{r}{{\small{}0.001}} & \multicolumn{1}{r}{{\small{}0.001}} & \multicolumn{1}{r}{{\small{}0.001}} & \multicolumn{1}{r}{{\small{}0.001}} & \multicolumn{1}{r}{{\small{}0.001}} & \multicolumn{1}{r}{{\small{}0.001}} & \multicolumn{1}{r}{{\small{}0.001}} & \multicolumn{1}{r}{{\small{}0.001}} & \multicolumn{1}{r}{{\small{}0.989}}\tabularnewline
\hline
\end{tabular}
\par\end{centering}
\centering{}{\small{}Notes: }$a=10^{-4}${\small{}; }$\mu_{n,s}^{2}=1000\pi_{s}^{2}/\sigma_{v,s}^{2}$.
\end{table}

\begin{table}[H]

\caption{\label{tab:EstiHi} Estimation results for $S=10$, high endogeneity}

\begin{centering}
\begin{tabular}{cccccccc}
\hline
$F$ & $F_{eff}$ & $F_{r}$ & $\widehat{\beta}_{OLS}$ & $\widehat{\beta}_{2sls}$ & $\widehat{\beta}_{gmmf}$ & $W_{2sls}$ & $W_{gmmf}$\tabularnewline
\hline
\multicolumn{1}{r}{0.994} & \multicolumn{1}{r}{0.994} & \multicolumn{1}{r}{80.12} & \multicolumn{1}{r}{0.754} & \multicolumn{1}{r}{0.749} & \multicolumn{1}{r}{0.007} & \multicolumn{1}{r}{1.000} & \multicolumn{1}{r}{0.067}\tabularnewline
\multicolumn{1}{r}{} & \multicolumn{1}{r}{} & \multicolumn{1}{r}{} & \multicolumn{1}{r}{{\small{}(0.000)}} & \multicolumn{1}{r}{{\small{}(0.023)}} & \multicolumn{1}{r}{{\small{}(0.029)}} & \multicolumn{1}{r}{} & \multicolumn{1}{r}{}\tabularnewline
\hline
\end{tabular}
\par\end{centering}
\centering{}{\small{}Notes: means and (st.dev.) of $10,000$\ replications.
Rej.freq. of robust Wald tests at 5\% level.}
\end{table}

\begin{figure}[H]
\begin{centering}
\includegraphics[scale=0.55]{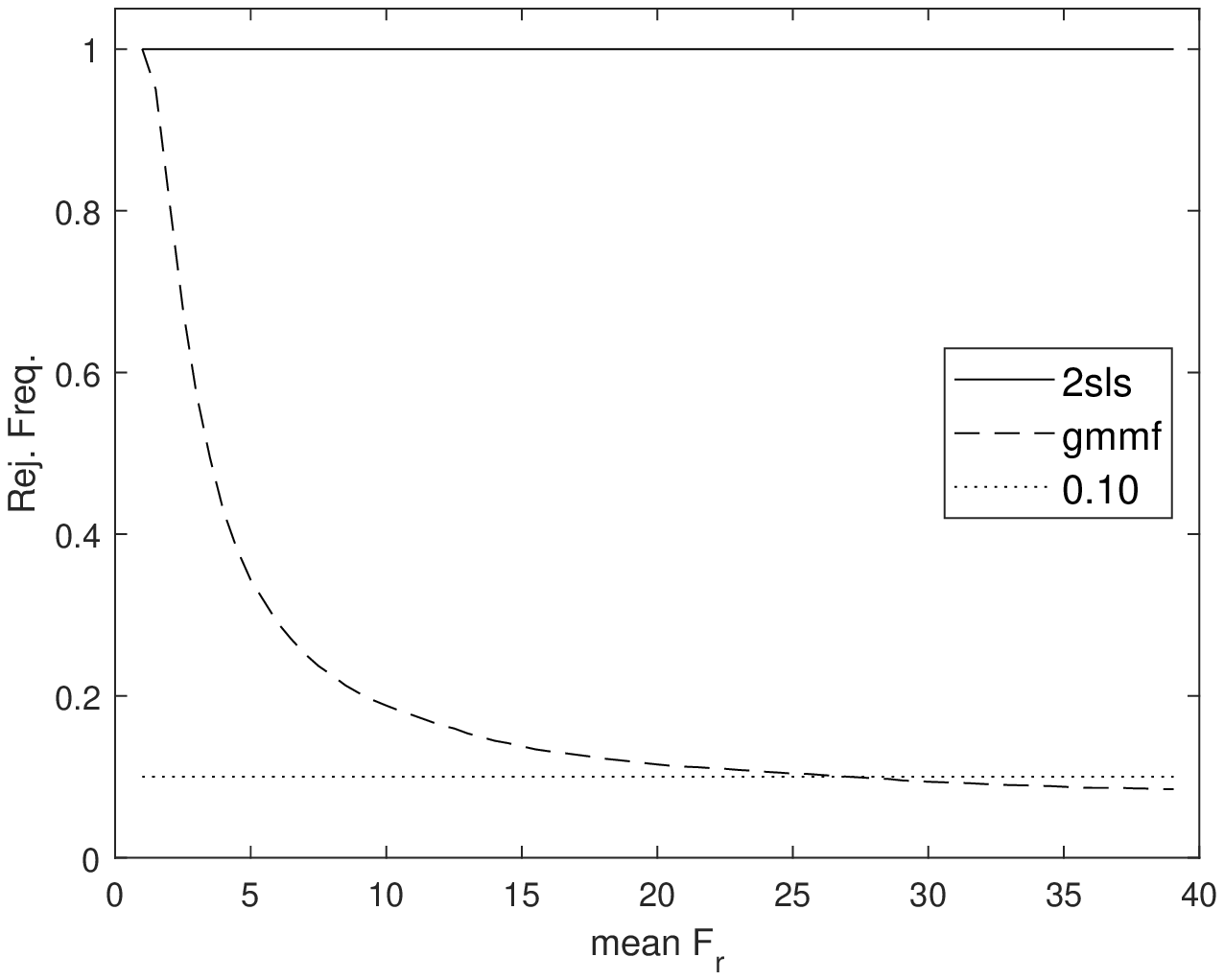}\includegraphics[scale=0.55]{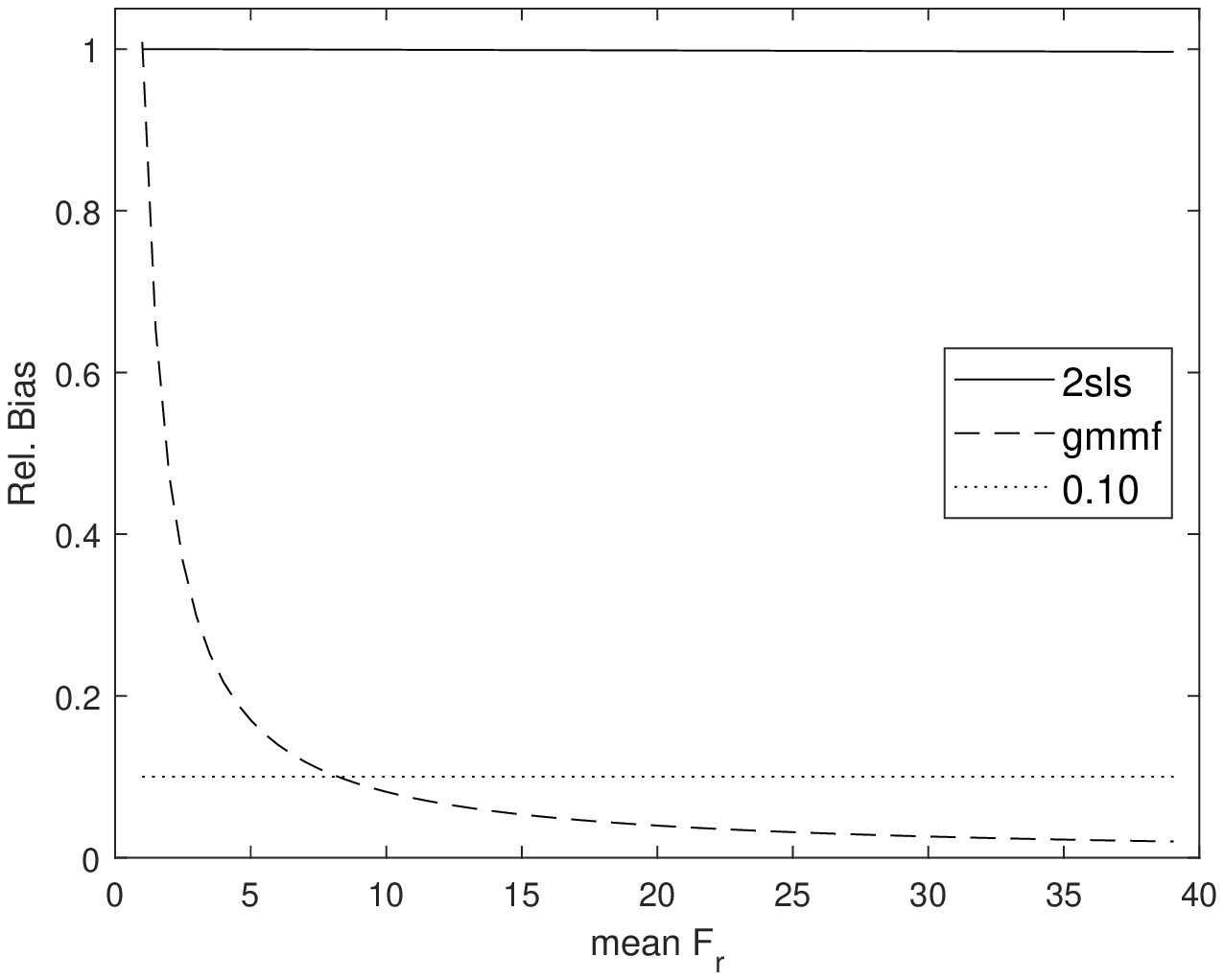}
\par\end{centering}
\caption{\label{fig:Rejfreqrelbhigh}Rejection frequencies of robust Wald tests
and relative bias, high endogeneity}

\end{figure}

Figure \ref{fig:Rejfreqrelbhigh} shows the rejection frequencies
of the robust Wald tests and the relative bias of the 2SLS and GMMf
estimators as a function of the value of the robust F-statistic, showing
a much better performance of the GMMf estimator.

\subsection{Testing for Weak Instruments}

\label{Sec:AppWIA}

Using the GMMf estimator as a generalization of the 2SLS estimator
to deal with general forms of first-stage heteroskedasticity, we investigate
here under what conditions the \citet{StockYogo2005} weak instruments
critical values derived for the non-robust F-test and the properties
of 2SLS estimator under full homoskedasticity apply to the robust
F-test and the properties of the GMMf estimator.

Consider again the standard linear model
\begin{eqnarray*}
y_{i} & = & x_{i}\beta+u_{i};\\
x_{i} & = & z_{i}^{\prime}\pi+v_{i},
\end{eqnarray*}
with conditional heteroskedasticity specified as
\begin{eqnarray*}
E\left[u_{i}^{2}|z_{i}\right] & = & \sigma_{u}^{2}\left(z_{i}\right);\,\,E\left[v_{i}^{2}|z_{i}\right]=\sigma_{v}^{2}\left(z_{i}\right);\,\,E\left[u_{i}v_{i}|z_{i}\right]=\sigma_{uv}\left(z_{i}\right),
\end{eqnarray*}
and, unconditionally,
\[
\sigma_{u}^{2}=E_{z}\left[\sigma_{u}^{2}\left(z_{i}\right)\right];\ \ \sigma_{v}^{2}=E_{z}\left[\sigma_{v}^{2}\left(z_{i}\right)\right];\ \ \sigma_{uv}=E_{z}\left[\sigma_{uv}\left(z_{i}\right)\right].
\]
Further, let
\[
\Omega_{u}=E\left[\sigma_{u}^{2}\left(z_{i}\right)z_{i}z_{i}^{\prime}\right];\ \ \Omega_{v}=E\left[\sigma_{v}^{2}\left(z_{i}\right)z_{i}z_{i}^{\prime}\right];\ \ \Omega_{uv}=E\left[\sigma_{uv}\left(z_{i}\right)z_{i}z_{i}^{\prime}\right],
\]
and assume that

\[
\left(\begin{array}{c}
\frac{1}{\sqrt{n}}Z^{\prime}u\\
\frac{1}{\sqrt{n}}Z^{\prime}v
\end{array}\right)\overset{d}{\rightarrow}\left(\begin{array}{c}
\psi_{zu}\\
\psi_{zv}
\end{array}\right)\sim N\left(\left(\begin{array}{c}
0\\
0
\end{array}\right),\left(\begin{array}{cc}
\Omega_{u} & \Omega_{uv}\\
\Omega_{uv}^{\prime} & \Omega_{v}
\end{array}\right)\right)
\]
\[
\left(\begin{array}{c}
\Omega_{u}^{-1/2}\frac{1}{\sqrt{n}}Z^{\prime}u\\
\Omega_{v}^{-1/2}\frac{1}{\sqrt{n}}Z^{\prime}v
\end{array}\right)\overset{d}{\rightarrow}\left(\begin{array}{c}
z_{u}\\
z_{v}
\end{array}\right)\sim N\left(\left(\begin{array}{c}
0\\
0
\end{array}\right),\left(\begin{array}{cc}
I_{k_{z}} & R\\
R^{\prime} & I_{k_{z}}
\end{array}\right)\right),
\]
where
\[
R=\Omega_{u}^{-1/2}\Omega_{uv}\Omega_{v}^{-1/2}.
\]

Let $\widehat{\pi}=\left(Z^{\prime}Z\right)^{-1}Z^{\prime}x$ be the
OLS estimator of $\pi$ and $\widehat{v}=x-Z\widehat{\pi}$ the OLS
residual. For the GMMf estimator, we have

\[
\widehat{\beta}_{gmmf}=\beta+\left(x^{\prime}Z\widehat{\Omega}_{v}^{-1}Z^{\prime}x\right)^{-1}x^{\prime}Z\widehat{\Omega}_{v}^{-1}Z^{\prime}u
\]
where, as before, $\widehat{\Omega}_{v}=\sum_{i=1}^{n}\widehat{v}_{i}^{2}z_{i}z_{i}^{\prime}$.
Assume that conditions are such that
\[
\frac{1}{n}Z^{\prime}Z\overset{p}{\rightarrow}E\left[z_{i}z_{i}^{\prime}\right]=Q_{zz}
\]
\[
\frac{1}{n}\widehat{\Omega}_{v}\overset{p}{\rightarrow}\Omega_{v}=E\left[\sigma_{v}^{2}\left(z_{i}\right)z_{i}z_{i}^{\prime}\right].
\]

For weak instrument asymptotics, let
\[
\pi=\frac{c}{\sqrt{n}},
\]
then
\begin{eqnarray*}
x^{\prime}Z\widehat{\Omega}_{v}^{-1}Z^{\prime}x & = & \left(Z\frac{c}{\sqrt{n}}+v\right)^{\prime}Z\widehat{\Omega}_{v}^{-1}Z^{\prime}\left(Z\frac{c}{\sqrt{n}}+v\right)\\
 & = & \frac{1}{n}c^{\prime}Z^{\prime}Z\widehat{\Omega}_{v}^{-1}Z^{\prime}Zc+\frac{2}{\sqrt{n}}c^{\prime}Z^{\prime}Z\widehat{\Omega}_{v}^{-1}Z^{\prime}v+v^{\prime}Z\widehat{\Omega}_{v}^{-1}Z^{\prime}v\\
 &  & \overset{d}{\rightarrow}\left(\lambda+z_{v}\right)^{\prime}\left(\lambda+z_{v}\right)
\end{eqnarray*}
where
\[
\lambda=\Omega_{v}^{-1/2}Q_{zz}c.
\]
It follows that
\begin{equation}
F_{r}=\frac{1}{k_{z}}x^{\prime}Z\widehat{\Omega}_{v}^{-1}Z^{\prime}x\overset{d}{\rightarrow}\chi_{k_{z}}^{2}\left(\lambda^{\prime}\lambda\right)/k_{z}.\label{Frwia}
\end{equation}
For the numerator, we have
\begin{eqnarray*}
x^{\prime}Z\widehat{\Omega}_{v}^{-1}Z^{\prime}u & = & \left(Z\frac{c}{\sqrt{n}}+v\right)^{\prime}Z\widehat{\Omega}_{v}^{-1}Z^{\prime}u\\
 & = & \frac{2}{\sqrt{n}}c^{\prime}Z^{\prime}Z\widehat{\Omega}_{v}^{-1}Z^{\prime}u+v^{\prime}Z\widehat{\Omega}_{v}^{-1}Z^{\prime}u\\
 &  & \overset{d}{\rightarrow}\left(\lambda+z_{v}\right)^{\prime}\Omega_{v}^{-1/2}\Omega_{u}^{1/2}z_{u}.
\end{eqnarray*}
For the OLS estimator,
\[
\widehat{\beta}_{ols}-\beta=\frac{x^{\prime}u}{x^{\prime}x}=\frac{\frac{c^{\prime}}{\sqrt{n}}Z^{\prime}u+v^{\prime}u}{\frac{1}{n}c^{\prime}Z^{\prime}Zc+\frac{2}{\sqrt{n}}c^{\prime}Z^{\prime}v+v^{\prime}v}\overset{p}{\rightarrow}\frac{\sigma_{uv}}{\sigma_{v}^{2}}.
\]
As $E\left[z_{u}|z_{v}\right]=Rz_{v}$, it follows for the relative
bias that
\begin{equation}
\frac{E\left[\widehat{\beta}_{gmmf}-\beta\right]}{E\left[\widehat{\beta}_{ols}-\beta\right]}\rightarrow\frac{\sigma_{v}^{2}}{\sigma_{vu}}E\left[\frac{\left(\lambda+z_{v}\right)^{\prime}\Omega_{v}^{-1/2}\Omega_{u}^{1/2}Rz_{v}}{\left(\lambda+z_{v}\right)^{\prime}\left(\lambda+z_{v}\right)}\right]\label{relbf}
\end{equation}
\[
\ \ \ \ \ \ \ \ \ \ \ \ \ \ \ \ \ \ \ \ \ \ \ \ \ \ \ \ =\frac{\sigma_{v}^{2}}{\sigma_{uv}}E\left[\frac{\left(\lambda+z_{v}\right)^{\prime}\Omega_{v}^{-1/2}\Omega_{uv}\Omega_{v}^{-1/2}z_{v}}{\left(\lambda+z_{v}\right)^{\prime}\left(\lambda+z_{v}\right)}\right].
\]
Therefore it follows that if $\Omega_{uv}=\delta\Omega_{v}$ and $\sigma_{uv}=\delta\sigma_{v}^{2}$,
for an arbitrary constant $\delta$, then
\begin{equation}
\frac{E\left[\widehat{\beta}_{gmmf}-\beta\right]}{E\left[\widehat{\beta}_{ols}-\beta\right]}\rightarrow E\left[\frac{\left(\lambda+z_{v}\right)^{\prime}z_{v}}{\left(\lambda+z_{v}\right)^{\prime}\left(\lambda+z_{v}\right)}\right].\label{relbwia}
\end{equation}
The conditions $\Omega_{uv}=\delta\Omega_{v}$ and $\sigma_{uv}=\delta\sigma_{v}^{2}$
are satisfied if $\sigma_{uv}\left(z_{i}\right)=\delta\sigma_{v}^{2}\left(z_{i}\right)$
for all $z_{i}\in\mathcal{Z}$.

The results (\ref{Frwia}) and (\ref{relbwia}) are the same as the
\citet{StaigerStock1997} and \citet{StockYogo2005} results for the
2SLS estimator and full conditional homoskedasticity,
\[
\left(\begin{array}{c}
\psi_{zu}\\
\psi_{zv}
\end{array}\right)\sim N\left(\left(\begin{array}{c}
0\\
0
\end{array}\right),\left(\begin{array}{cc}
\sigma_{u}^{2} & \sigma_{uv}\\
\sigma_{uv} & \sigma_{v}^{2}
\end{array}\right)\otimes Q_{zz}\right)
\]
and with $\lambda=\sigma_{v}^{-1}Q_{zz}^{1/2}c$. Therefore the Stock
and Yogo (2005) critical values apply to the robust F-statistic and
relative bias (\ref{relbf}) of the GMMf estimator if $\Omega_{uv}=\delta\Omega_{v}$
and $\sigma_{uv}=\delta\sigma_{v}^{2}$. For the grouped data IV example,
this condition is fulfilled if $\sigma_{uv,s}=\delta\sigma_{v,s}^{2}$
for all $s=1,...,S$.

\bigskip{}

For the Wald test, we have
\[
V\widehat{a}r\left(\widehat{\beta}_{gmmf}\right)=\left(x^{\prime}Z\widehat{\Omega}_{v}^{-1}Z^{\prime}x\right)^{-1}x^{\prime}Z\widehat{\Omega}_{v}^{-1}\widehat{\Omega}_{u}\widehat{\Omega}_{v}^{-1}Z^{\prime}x\left(x^{\prime}Z\widehat{\Omega}_{v}^{-1}Z^{\prime}x\right)^{-1}
\]
and so
\[
W_{gmmf}\left(\beta\right)=\frac{\left(\widehat{\beta}_{gmmf}-\beta\right)^{2}\left(x^{\prime}Z\widehat{\Omega}_{v}^{-1}Z^{\prime}x\right)^{2}}{x^{\prime}Z\widehat{\Omega}_{v}^{-1}\widehat{\Omega}_{u}\widehat{\Omega}_{v}^{-1}Z^{\prime}x}=\frac{\left(x^{\prime}Z\widehat{\Omega}_{v}^{-1}Z^{\prime}u\right)^{2}}{x^{\prime}Z\widehat{\Omega}_{v}^{-1}\widehat{\Omega}_{u}\widehat{\Omega}_{v}^{-1}Z^{\prime}x}.
\]
Then,
\begin{eqnarray*}
\frac{1}{n}\widehat{\Omega}_{u} & = & \frac{1}{n}\sum_{i=1}^{n}\widehat{u}_{i}^{2}z_{i}z_{i}^{\prime}\\
 & = & \frac{1}{n}\sum_{i=1}^{n}\left(u_{i}-x_{i}\left(\widehat{\beta}_{gmmf}-\beta\right)\right)^{2}z_{i}z_{i}^{\prime}\\
 & = & \frac{1}{n}\sum_{i=1}^{n}\left(u_{i}^{2}-2u_{i}x_{i}\left(\widehat{\beta}_{gmmf}-\beta\right)+x_{i}^{2}\left(\widehat{\beta}_{gmmf}-\beta\right)^{2}\right)z_{i}z_{i}^{\prime}\\
 &  & \overset{d}{\rightarrow}\Omega_{u}-2\Omega_{uv}\left(\widehat{\beta}_{gmmf}-\beta\right)+\Omega_{v}\left(\widehat{\beta}_{gmmf}-\beta\right)^{2},
\end{eqnarray*}
and so
\begin{eqnarray*}
 &  & x^{\prime}Z\widehat{\Omega}_{v}^{-1}\widehat{\Omega}_{u}\widehat{\Omega}_{v}^{-1}Z^{\prime}x\overset{d}{\rightarrow}\left(\lambda+z_{v}\right)^{\prime}\Omega_{v}^{-1/2}\Omega_{u}\Omega_{v}^{-1/2}\left(\lambda+z_{v}\right)\\
 &  & -2\left(\lambda+z_{v}\right)^{\prime}\Omega_{v}^{-1/2}\Omega_{uv}\Omega_{v}^{-1/2}\left(\lambda+z_{v}\right)\left(\widehat{\beta}_{gmmf}-\beta\right)\\
 &  & +\left(\lambda+z_{v}\right)^{\prime}\left(\lambda+z_{v}\right)\left(\widehat{\beta}_{gmmf}-\beta\right)^{2}.
\end{eqnarray*}
This results in
\[
W_{gmmf}\overset{d}{\rightarrow}\frac{q_{2}^{2}}{a-2bq_{2}/\eta_{1}+q_{2}^{2}/\eta_{1}},
\]
where
\begin{eqnarray*}
q_{2} & = & \left(\lambda+z_{v}\right)^{\prime}\Omega_{v}^{-1/2}\Omega_{u}^{1/2}z_{u}\\
a & = & \left(\lambda+z_{v}\right)^{\prime}\Omega_{v}^{-1/2}\Omega_{u}\Omega_{v}^{-1/2}\left(\lambda+z_{v}\right)\\
b & = & \left(\lambda+z_{v}\right)^{\prime}\Omega_{v}^{-1/2}\Omega_{uv}\Omega_{v}^{-1/2}\left(\lambda+z_{v}\right)\\
\eta_{1} & = & \left(\lambda+z_{v}\right)^{\prime}\left(\lambda+z_{v}\right).
\end{eqnarray*}

If $\Omega_{u}=\kappa\Omega_{v}$ and $\Omega_{uv}=\delta\Omega_{v}$,
so $R=\frac{\delta}{\sqrt{\kappa}}I_{k_{z}}$, then
\begin{equation}
W_{gmmf}\left(\beta\right)\overset{d}{\rightarrow}\frac{\kappa\eta_{2}^{2}}{\kappa\eta_{1}-2\delta\sqrt{\kappa}\eta_{2}+\kappa\eta_{2}^{2}/\eta_{1}}=\frac{\eta_{2}^{2}/\eta_{1}}{1-2\rho\eta_{2}/\eta_{1}+\left(\eta_{2}/\eta_{1}\right)^{2}}\label{Waldwia}
\end{equation}
where
\begin{eqnarray*}
\eta_{2} & = & \left(\lambda+z_{v}\right)^{\prime}z_{u}\\
\rho & = & \frac{\delta}{\sqrt{\kappa}}.
\end{eqnarray*}
Conditions $\Omega_{u}=\kappa\Omega_{v}$ and $\Omega_{uv}=\delta\Omega_{v}$
are satisfied if $\sigma_{uv}\left(z_{i}\right)=\delta\sigma_{v}^{2}\left(z_{i}\right)$
and $\sigma_{u}^{2}\left(z_{i}\right)=\kappa\sigma_{v}^{2}\left(z_{i}\right)$
for all $z_{i}\in\mathcal{Z}$, and then $\rho=\frac{\sigma_{uv}}{\sigma_{u}\sigma_{v}}$.

Result (\ref{Waldwia}) is the same as that of the \citet{StaigerStock1997}
and \citet{StockYogo2005} result for the 2SLS based Wald test under
conditional homoskedasticity with the maximum size distortion at $\rho^{2}=1$.
Hence the \citet{StockYogo2005} Wald size based critical values apply
in the heteroskedastic case to the GMMf based Wald test if $\Omega_{u}=\kappa\Omega_{v}$
and $\Omega_{uv}=\delta\Omega_{v}$, with again the maximum size distortion
at $\delta^{2}/\kappa=1$.

\bibliographystyle{ecta}
\bibliography{RobF}

\end{document}